**Title:** Magnetostructural study of the $(Mn,Fe)_3(P,Si)$ system

**Authors:** J. V. Leitão*[1], You Xinmin[2], L. Caron[1,3], E. Brück[1]

(1) Fundamental Aspects of Materials and Energy, Faculty of Applied Sciences, TU Delft, Mekelweg 15, 2629 JB Delft, Netherlands
(2) Nanjing University of Technology, Nanjing, China
(3) Department of Engineering Sciences, Uppsala University, Box 534, 751 21 Uppsala, Sweden
* Corresponding author: J.C.Vieiraleitao@tudelft.nl

**Abstract:** Using X-ray diffraction, DSC and magnetization measurements, a magnoestructural map of the $(Mn,Fe)_3(Si,P)$ system was assembled and reported in the current paper. Besides the already known cubic phase for $Mn_{3-x}Fe_xSi$ system and the tetragonal and orthorhombic phases for the $Mn_{3-x}Fe_xP$ system, a novel hexagonal phase has been observed for $Mn_{3-x}Fe_xSi_{1-y}P_y$, within the approximate range of $0.2<x<2.0$ and $0.2<y<0.9$. Magnetization measurements both confirm and further detail the already known properties of the $Mn_{3-x}Fe_xSi$ and $Mn_{3-x}Fe_xP$ systems.

**Keywords:** Magnetostructural map, $Mn_3Si$, $Mn_3P$, $Fe_3Si$, $Fe_3P$, $Mn_{3-x}Fe_xSi$, $Mn_{3-x}Fe_xP$, magnetocaloric effect

## 1 – Introduction

Having the initial motivation to study the potential of the $(Mn,Fe)_3(P,Si)$ system for magnetic cooling applications using the magnetocaloric effect (MCE), the current paper reports the curious and unique magnetostructural properties of this system.
The MCE represents one of the best chances of creating a more efficient refrigeration system, alternative to the usual vapor compression technology. This effect, although known since the late 1920's [1-3], and its possible application in a magnetic cooling device since 1976 [4], has still to be effectively applied to a completely viable commercial room temperature cooling device. As such, there has been a constant search for more effective, cheap and non-toxic working materials that may open the possibility for such a refrigerator.
In this context, attention has been paid to the use of transition metal alloys as a possible cheap and non-toxic alternative to other materials with already proven magnetocaloric properties, such as the $Gd_5(Si_xGe_{1-x})_4$[5] or $MnAs_{1-x}Sb_x$[6] systems.

As the magnetocaloric effect is directly related to the change in magnetic entropy, as a consequence of transitions between states of magnetic order and disorder, it will be maximized around large and sharp jumps in magnetization, such as those present around the Curie Temperature ($T_C$), or other magnetic phase transitions; also, it may be described as a function of the area between magnetic isotherms [7].
Under this perspective, even thought several compositional areas of the studied system presented significant magnetization shifts, none of these were sufficiently large or sharp to make this system appropriate for magnetic cooling purposes. Still, the amount of information and results gathered from this study has made it possible to outline the basic properties and assemble a magnetostructural phase diagram for the $(Mn,Fe)_3(P,Si)$ system.

## 2 – Material overview

### 2.1 – The $Mn_{3-x}Fe_xSi$ system

The $Mn_{3-x}Fe_xSi$ system has been studied both for its properties as an itinerant-electron antiferromagnet and as a half-metallic ferromagnet (HMF), at different compositional ranges. In its antiferromagnetic phase (Mn rich) it has been studied in the context of spin-wave excitation by means of neutron inelastic scattering [8]. As a HMF (Fe rich), having a band gap in one spin at the Fermi level whereas the other spin is strongly metallic, which results in a complete spin polarization of the conduction electrons at the Fermi level, it is of interest in the field of spintronics [9].

The $Mn_3Si$ compound crystallizes in the cubic Fm3m structure ($AlFe_3$ prototype) [8,10,11]. It is a bcc structure with two distinct Mn atom sites, A (MnI) and B/B' (MnII) [8,10], as shown in Figure 1a. It is reported to have magnetic moments of 0.19 $\mu_B$, and 1.7 $\mu_B$ [8,10-12] respectively. This compound orders antiferromagnetically below a Néel Temperature ($T_N$) of about 25 K, being paramagnetic above this temperature [8,10-12].

$Fe_3Si$ crystallizes in the same cubic Fm3m structure as $Mn_3Si$, the Fe atoms on the A site have a magnetic moment of 2.4 $\mu_B$ while those on the B/B' site exhibit a magnetic moment of 1.2 $\mu_B$ [13,14]. The moments are ordered ferromagnetically at room temperature with a $T_C$ of about 840 K [15,16].

The whole of the $Mn_{3-x}Fe_xSi$ system maintains the same stable Fm3m crystal structure for $0<x<3$. A minor secondary phase, either tetragonal or hexagonal, has also been reported by S. Yoon et al. for Mn contents between 1.80 and 3 [17].

From the $Fe_3Si$ compound, with increasing Mn content, $T_C$ drops linearly from 840 K to approximately 300 K at the composition $Mn_{0.7}Fe_{2.3}Si$ [16,17]. From this composition on it is also found, for temperatures around 50 K, a transition between ferromagnetic and antiferromagnetic states [18], what H. Miki et al. described as canted ferromagnetism [19]. This second transition eventually fades away at the composition of $Mn_{1.8}Fe_{1.2}Si$. In this interval $T_C$ decreases at a much smaller rate, reaching a value of 65 K at the above mentioned composition. Further addition of Mn from this point triggers an antiferromagnetic behavior in this system. From this content on, $T_N$ drops smoothly from the last value of $T_C$ until it reaches 25 K at $Mn_3Si$ [19].

Regarding site preferences of the Fe and Mn atoms, in the parent compound $Fe_3Si$, the addition of Mn atoms, for compositions between 0 and 0.75, preferably occupies the A sites. Within this interval, $T_C$ continuously decreases from 840 K to 370 K, while the magnetic moment of this site remains relatively constant, meaning that Fe(A) and Mn(A) atoms exhibit approximately the same moment [18]. For compositions above 0.75 the Mn atoms begin to occupy half of the B sites, so that the B sublattice can be split into B and B' sites [20], although the A sites are only found to be completely filled for a Mn content above 1.5 [18].

For Mn composition between 0.75 and 1.8, we then find the already mentioned canted ferromagnetism displaying a transition between canted ferromagnetic and antiferromagnetic states [18,19].

## 2.2 – The $Mn_{3-x}Fe_xP$ system

As far as practical application goes, this system has been studied in the context of nuclear-reactor material research, mostly the $Fe_3P$ compound [21]. Beyond this it does not seem to provoke much interest outside the academic sphere.
Both the $Mn_3P$ and $Fe_3P$ compounds crystallize in the tetragonal $I\bar{4}$ structure [22,23] ($Ni_3P$ prototype), Figure 1b, with three different Mn/Fe sites [23]. $Mn_3P$ is an antiferromagnet with a $T_N$ of about 115 K and a magnetic moment of 1.69 $\mu_B$ per Mn atom [24]. $Fe_3P$ on the other hand is a ferromagnet with a $T_C$ of about 700 K and a magnetic moment of 1.89 $\mu_B$ per Fe atom [24].

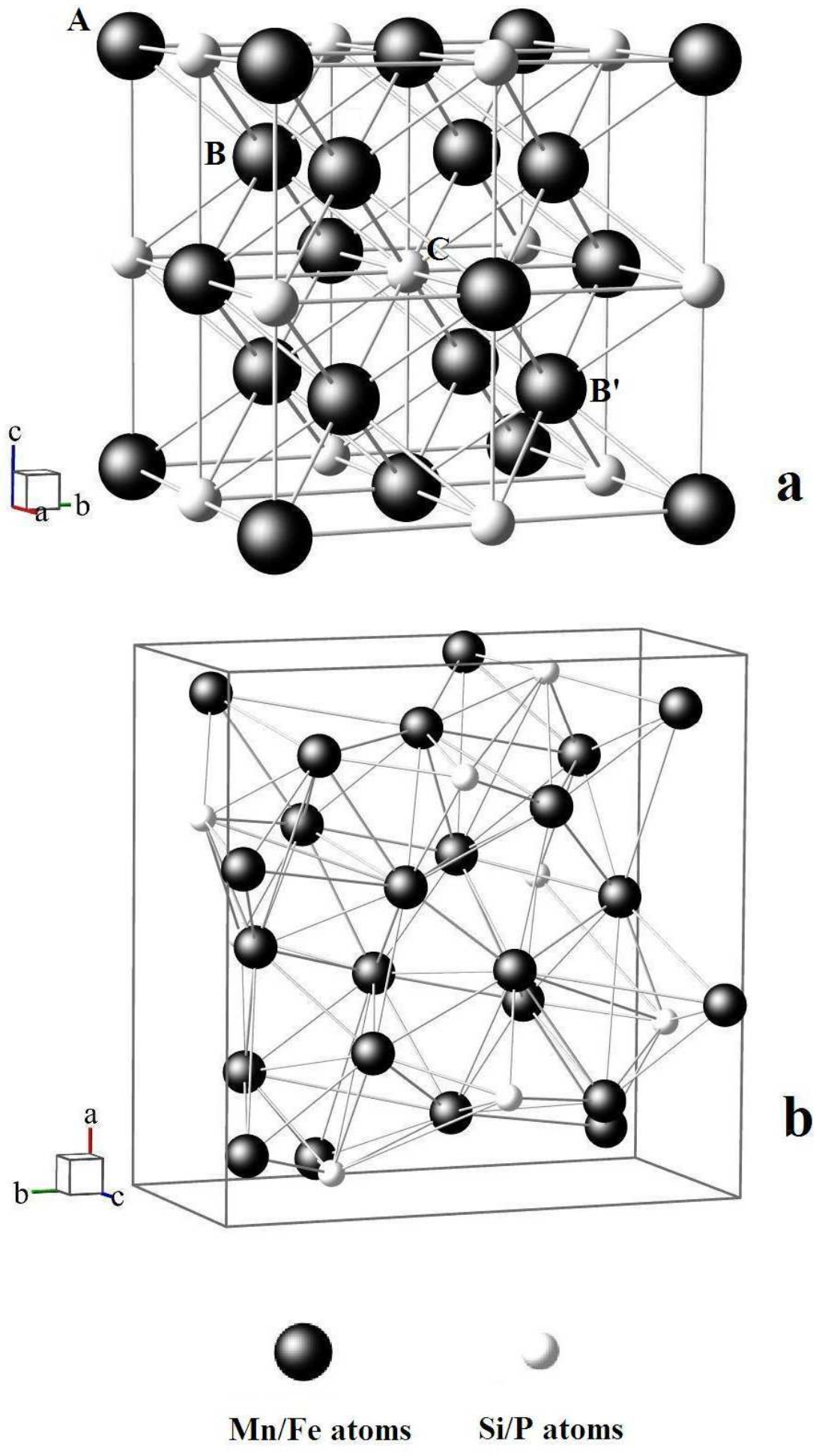

**Figure 1 - a) The $Mn_3Si$ Fm3m cubic structure, showing the MnI atoms on the A sites, the MnII atoms on the B/B' sites and the Si atoms on the C sites; b) The $Mn_3P$ and $Fe_3P$ $I\bar{4}$ tetragonal structure**

The $Mn_{3-x}Fe_xP$ system maintains the same $I\bar{4}$ tetragonal crystal structure for 0<x<1 [25,26] and again for 2.2<x<3 [24,26], approximately. In the gap between these two intervals an orthorhombic structure is reported [26]. In this structure the moments order antiferromagnetic with a relatively constant $T_N$ of about 340 K [26].
Mitita et al. [26], having studied the magnetic properties of $Fe_{3-x}M_xP$, with M=Cr, Mn, Co, Ni, report that this orthorhombic structure only occurred in the Mn compounds, all the other systems maintained the $I\bar{4}$ tetragonal structure along the studied compositions.

## 3 - Experimental Procedure and Characterization

### 3.1 – Sample preparation

All samples produced belonging to the $(Mn,Fe)_3(Si,P)$ system were prepared from the appropriate amounts of 99+% iron powder, 99.99% granular silicon, 99% red phosphorous powder and 99.9% manganese chips reduced at 600 ºC under a hydrogen atmosphere in order to remove oxides. The samples were milled in a Fritsch Pulverisette planetary mill for 6 hours (3 hours with 5 minute breaks every 5 minutes to prevent overheating) at 360 rpm in 80 ml hardened steel crucibles, each containing fifteen 4 g hardened steel balls, amounting to a sample\ball ratio of 0.083(3) with the sample mass (5 g).
The samples were then compacted into 10 mm pellets with a pressure of 150 $kgf/cm^2$ and sealed into quartz tubes with an atmosphere of 200 mbar of argon. Finally these were annealed in a vertical resistive furnace for 100 hours at 950 ºC and quenched in water at room temperature.
Such a procedure was selected so as to make it appropriate for all possible composition variations of the $(Mn,Fe)_3(Si,P)$ system. The use of an arc melting furnace (a common procedure for the production of samples belonging to the $Mn_{3-x}Fe_xSi$ system [15,17,27]) was found to be unsuitable for the production of samples containing Phosphorous.

### 3.2 – Characterization methods

In order to check the homogeneity and crystal structure of our samples room temperature X-ray diffraction was performed in an X'Pert PRO X-ray diffractometer with Cu $K_\alpha$ radiation from PANalytical. The resulting diffraction patterns were analyzed using the software X'Pert HightScore and FullProf's implementation of the Rietveld refinement method [28].
Magnetic measurements were performed in two different magnetometers, equipped with superconducting quantum interference devices (SQUID), a MPMS-5S and a MPMS XL model, both from Quantum Design.
The measurements taken were temperature sweeps from 5 K to either 370 K (MPMS-5S) or 400 K (MPMS XL) with a fixed applied magnetic field.
Further DSC (Differential scanning calorimetry) measurements were performed on those samples whose transition temperatures exceeded the temperature range of our magnetometers.

A Q2000 model from TA Instruments-Waters LLC was used for this end, performing temperature sweeps from 0 to 500 ºC at a rate of 20 ºC per minute.

## 4 – Results and discussion

### 4.1 – Structural Results

All the structural information regarding the $Mn_{3-x}Fe_xSi$ and $Mn_{3-x}Fe_xP$ systems mentioned in section 2 was confirmed.
The Fm3m cubic structure of the $Mn_3Si$ and $Fe_3Si$ was found to exist up to the Phosphorous substitution of $Mn_3Si_{0.8}P_{0.2}$ on the Mn rich side of the diagram and up to $Fe_3Si_{0.4}P_{0.6}$ on the Fe rich side. The secondary tetragonal or hexagonal phases reported by S. Yoon at al. [17] in cubic samples with Mn content above 1.80 were not observed. Such a result may be a direct consequence of the different sample preparation procedures used for the current paper and those used by the mentioned authors, namely, arc melting, annealing for 24 hours at 800 ºC, and quenching into cold water.
The $I\bar{4}$ tetragonal phase of the $Mn_3P$ and $Fe_3P$, on the Mn rich side, was found to begin immediately after de Fm3m phase, around the composition of $Mn_3Si_{0.8}P_{0.2}$. On the Fe rich side this structure was found to exist up to the Si substitution of $Fe_3Si_{0.4}P_{0.6}$.
The diffraction data from samples displaying the orthorhombic phase was found to be consistent with the Pmmm space group, Fig. 2a. This structure was also found to exist up to the composition of $MnFe_2Si_{0.5}P_{0.5}$.
A fourth crystal structure, consistent with the hexagonal P6/mmm space group, was found for values in the range of $0.2<x<1.95$ and $0.2<y<0.9$, for $Mn_{3-x}Fe_xSi_{1-y}P_y$, Fig. 2b. It is possible that this structure is the same as the one observed by S. Yoon at al. [17] as a secondary phase in cubic samples with Mn content above 1.80, but, as these authors did not provide any additional information regarding this secondary phase we cannot confirm this hypothesis. As such, the detection of this structure in this system can be said to be a complete novelty and to have never been reported in literature.
The lattice constant *a* for the cubic phase has been found to decrease with Fe content, going from 5.72 Å for $Mn_3Si$ to 5.65 Å for $Fe_3Si$, in excellent accordance with the values found by S. Yoon at al. [17]. However, the substitution of P seems to promote an increases in the lattice parameter *a*, as it was found both for $Fe_3Si_{0.8}P_{0.2}$ and $Fe_3Si_{0.5}P_{0.5}$ a value for *a* of 5.66 Å.
The *a* and *c* constants in the Mn rich tetragonal phase also present a good accordance to the literature [25, 26], decreasing with increasing Fe content. Si substitution on this area also increases both the *a* and *c*, coming from an *a* of 9.179 Å and a *c* of 4.568 Å for $Mn_3P$ to 9.183 Å and 4.607 Å for $Mn_3Si_{0.2}P_{0.8}$.
On the Fe rich tetragonal phase, values of 9.104 Å and 4.4631 Å for *a* and *c*, respectively, were found for the $Fe_3P$ compound, much in accordance with the literature [24, 26]. We were unable to establish with certainty the evolution of the lattice parameter with Mn substitution in this compositional area due to the presence of an orthorhombic secondary phase in our samples, preventing us from achieving a good fit of our X-ray diffraction patterns. However, it has been determined that Si substitution increases rather markedly both lattice parameter, having the composition of $Fe_3Si_{0.2}P_{0.8}$ an *a* of 9.1168 Å and a *c* of 4.4782 Å.

The lattice parameters of the orthorhombic phase were found to be slightly smaller than those reported by Mitita Goto et al. [26], being *a* = 8.9456 Å, *b* = 8.0079 Å and *c* = 4.3368 Å for the $Mn_{1.5}Fe_{1.5}P$ compound. Due to the secondary phases, belonging to both the Mn and the Fe rich tetragonal phase and the hexagonal phase, we were able to make a good fitting of our other orthorhombic samples.

The hexagonal phase, for the $Mn_{2.5}Fe_{0.5}Si_{0.5}P_{0.5}$, presents an *a* of 8.83 Å and a *c* of 10.89 Å. The values decrease with increasing Fe content, arriving at an *a* of 8.77 Å and a *c* of 10.64 Å for $Mn_{1.4}Fe_{1.6}Si_{0.5}P_{0.5}$. X-ray diffraction fitting for samples with a higher Fe content than 1.6 was found to be unreliable as these samples presented a larger amount of secondary phases belonging to the cubic and orthorhombic structures. The same problem was found with our samples with other Si\P ratios than the presented 0.5/0.5.

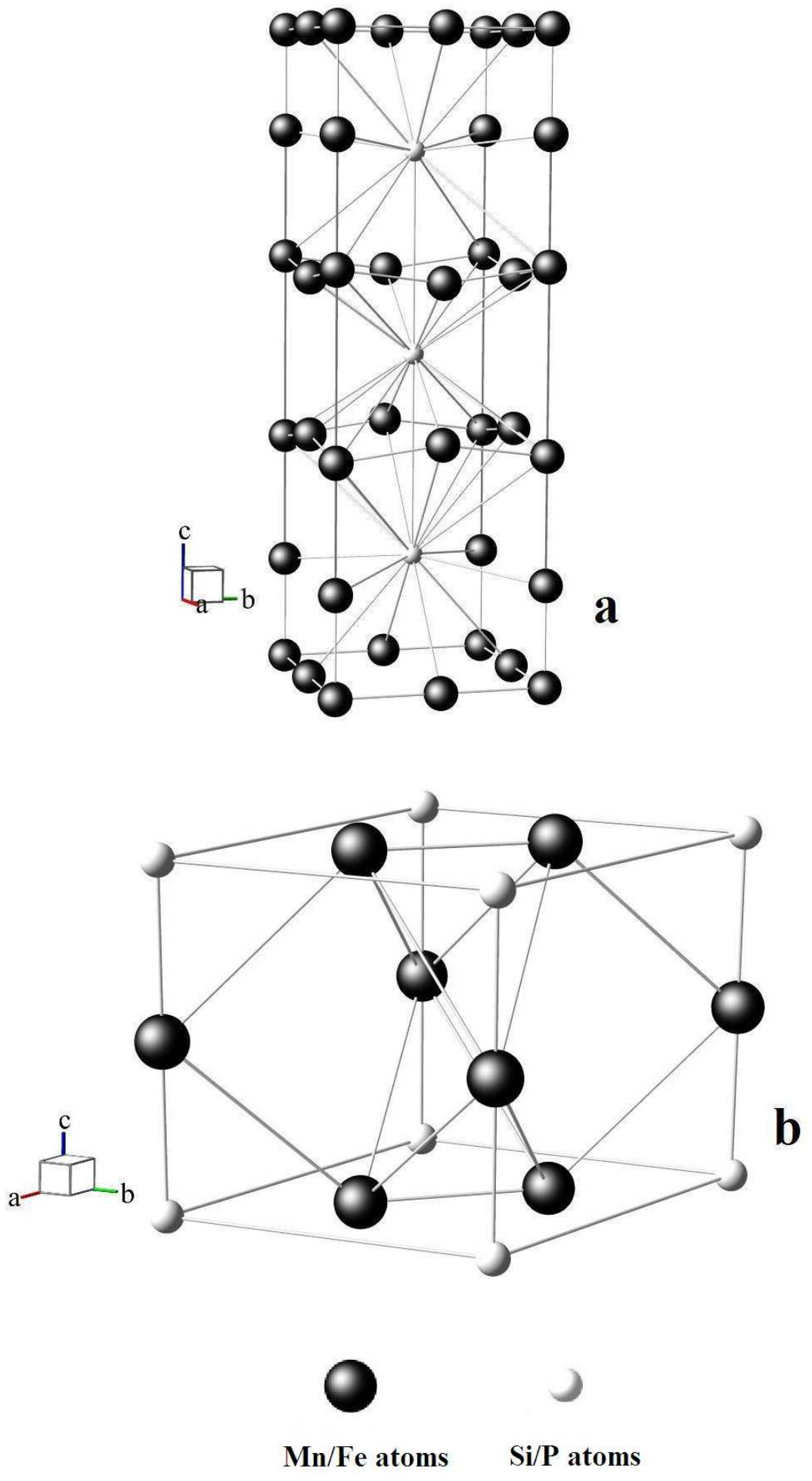

**Figure 2 - a) Pmmm orthorhombic structure of the $Mn_{3-x}Fe_xP$ (for values of x betewn 1 and 2, aproximatly); b) P6/mmm hexagonal structure of the $Mn_{3-x}Fe_xSi_{1-y}P_y$ (for 0.2<x<2.0 and 0.2<y<0.9)**

The borders between the various structural phases in this system were found to be significantly influenced by the annealing temperature used in their production. Having tested various annealing temperatures on a set of $MnFe_2Si_{0.5}P_{0.5}$ samples (displaying cubic, hexagonal and orthorhombic phases, see Table 1) it was found that the hexagonal phase increases with increasing annealing temperature, in detriment of the cubic phase. The orthorhombic phase also appears to increase with temperature but not significantly.

## 4.2 – Magnetic results

All of the samples belonging to the Pmmm orthorhombic structure, in accordance with the results from Mitita et al. [26], were found to be consistently antiferromagnetic.

The novel Hexagonal phase, similarly to the Cubic phase, was found to exhibit a Ferromagnetic-Paramagnetic transition above an Fe content of 1.2. This transition, like those present in cubic and tetragonal phases of this system, also presents the characteristics of a second order phase transition.

On all border regions samples were found to always exhibit multiple phases, as in fact rigid borders or discontinuities between different phases aren't observed. What was found were relatively wide bands where the bordering phases coexist, as such, border lines can only be estimated through the analysis of phase fractions on each sample. However, by taking into consideration magnetic measurements, a much clearer monitoring of this border may be obtained.

The border between the orthorhombic and tetragonal phases was mapped through the analysis of Si content in the magnetic behavior of the sample, as there is a very clear difference between the antiferromagnetic orthorhombic phase and the ferromagnetic tetragonal phase (Figure 3).

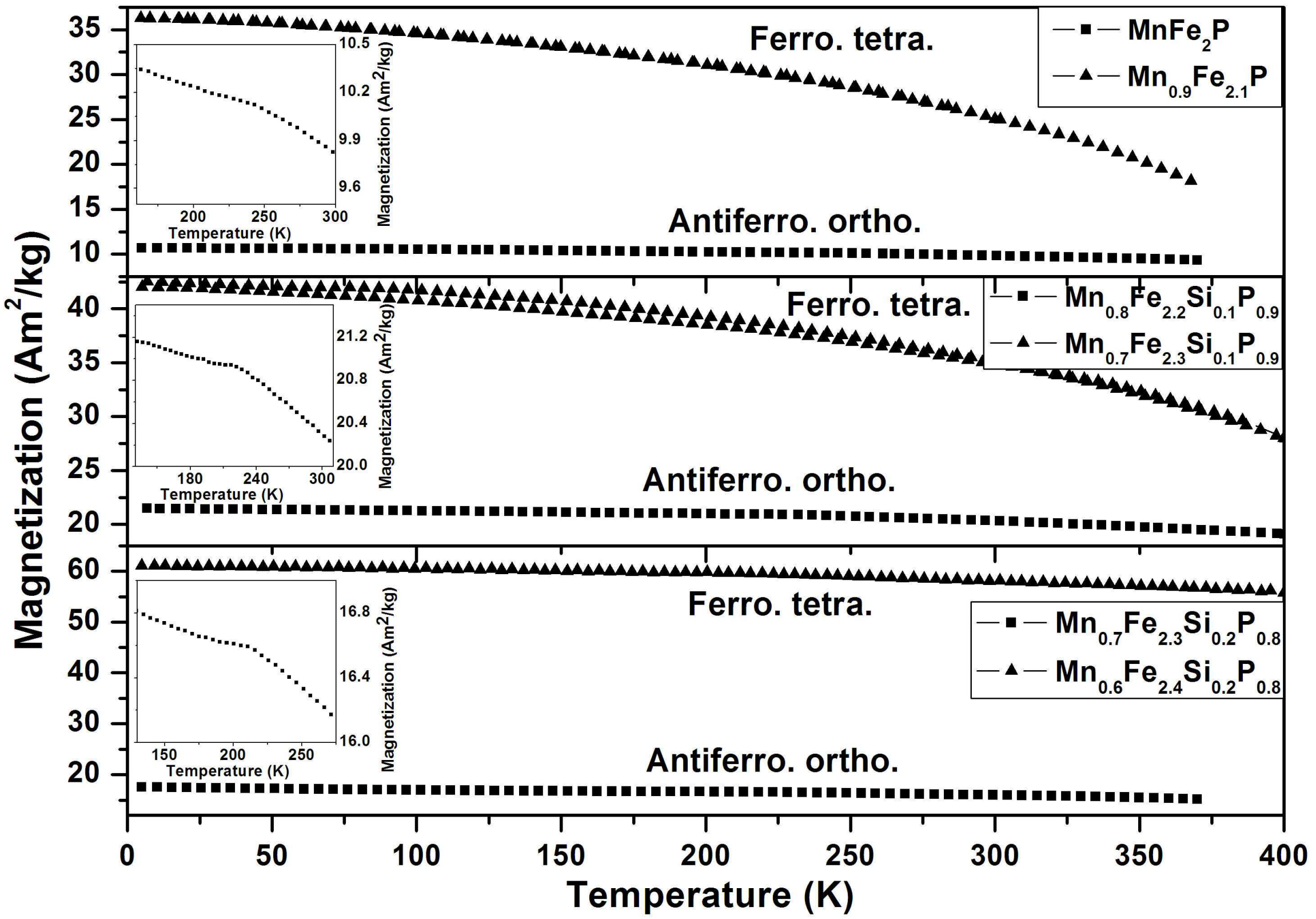

**Figure 3 - Magnetization versus temperature plots revealing the influence of Si content on the magnetic behaviors of the compositions on both sides of the border between the orthorhombic and tetragonal phases, taken with 1 Tesla of applied magnetic field. Inserts: detail on the magnetization curve of the aniferromagnetic orthorhombic samples, revealing the characteristic bump of an antiferromagnetic transition.**

Furthermore, assuming that the magnetization of the ferromagnetic tetragonal phase (Fe rich) decreases linearly with Mn content, through extrapolation of the measured values of magnetization at 5 K we can then estimate the percentage of the different phases present in our samples.
At 5 K we found for our $Mn_{0.5}Fe_{2.5}Si_{0.2}P_{0.8}$ and $Mn_{0.6}Fe_{2.4}Si_{0.2}P_{0.8}$ samples a magnetization of 72.78 $Am^2/kg$ and 61.19 $Am^2/kg$ respectively. Should this behavior continue linearly we would expect a value of around 49.6 $Am^2/kg$ for $Mn_{0.7}Fe_{2.3}Si_{0.2}P_{0.8}$, where instead we find the value of 17.52 $Am^2/kg$, showing that this sample is now predominantly antiferromagentic. Through simples calculus we can then assume that our $Mn_{0.7}Fe_{2.3}Si_{0.2}P_{0.8}$ sample is made up of about 35.4% of ferromagnetic tetragonal phase and 64.6% antiferromagnetic orthorhombic phase.
This method of phase percentage estimation, however useful, can only be used on border areas between structures with distinct magnetic behavior.

The temperature-induced transition from antiferromagnetism to canted ferromagnetism in compounds on the Si rich side of the structure diagram has been found to be completely independent of any compositional or magnetic field change, maintaining a relatively constant temperature of about 50 K between the Mn content of 1.8 and 1. Below this composition it rapidly decreases until it is no longer observable at a Mn content of 0.6, a result also observed by S. Yoon at al. [16,17]. The addition of P in this area has proven to reduce the overall magnetic moment of the sample and widen the already broad second order phase transition between the ferromagnetic and paramagnetic phases, meaning, an increase in $T_C$, as seen on Figure 4. This effect is probably due to the increase in the lattice parameters, and consequent atomic distance, that P substitution promotes in this structure.

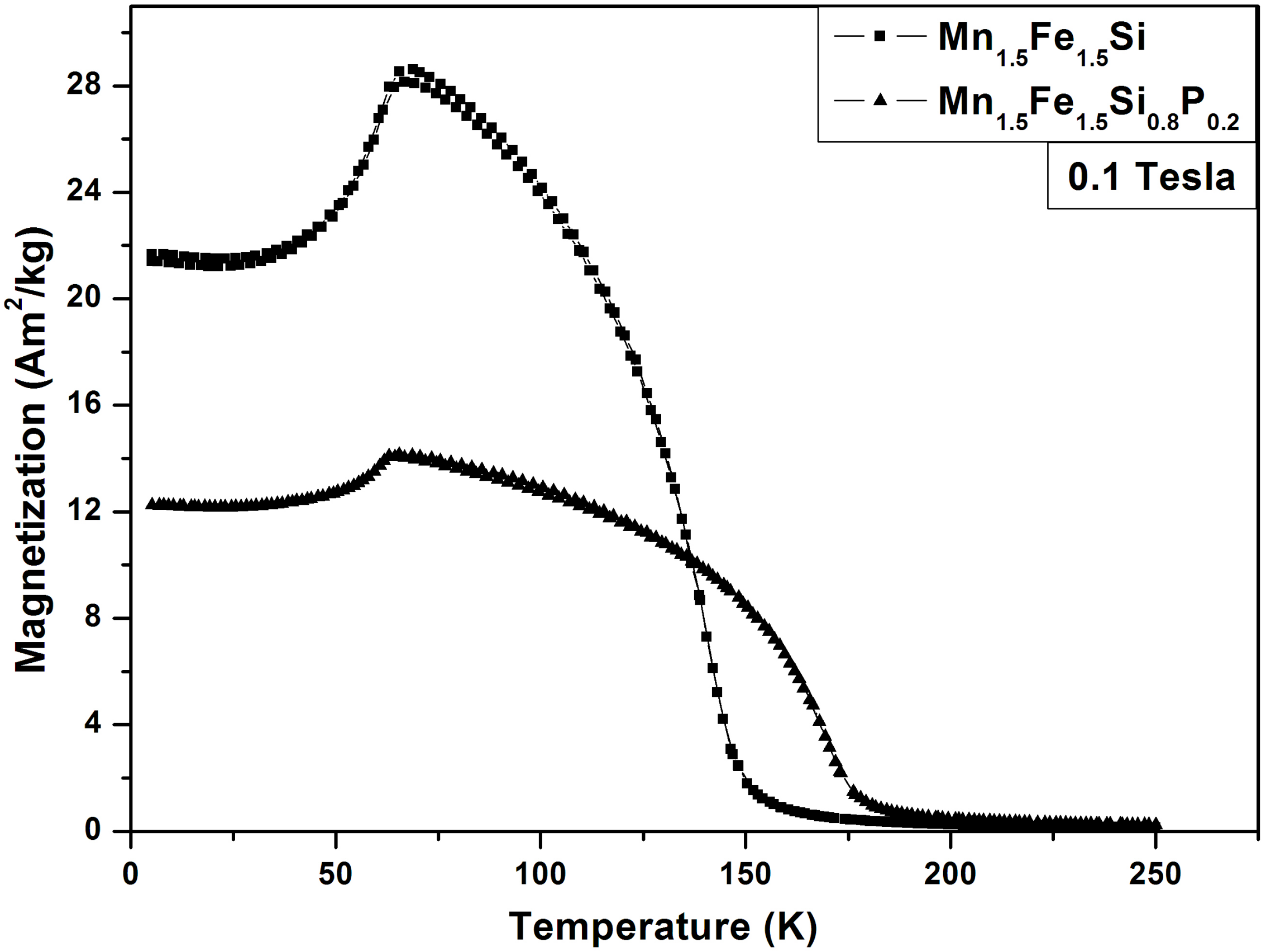

**Figure 4 - Magnetization versus temperature plot reveling the influence of P addition to the canted ferromagnetic phase of the (Mn,Fe)$_3$(Si,P) system**

Furthermore, an analysis of the temperature dependence of the magnetization for $Mn_{3-x}Fe_xSi$ samples, with 2<x<2.5, revealed an unusual magnetic behavior. Instead of observing a continuous increase in magnetization with increasing Fe content, abrupt jumps in magnetization were clearly observed, Figure 5. This result can be partially explained by the Mn and Fe atom site preferences described in section 2.1, although this hypothesis does not fully explain such a result.

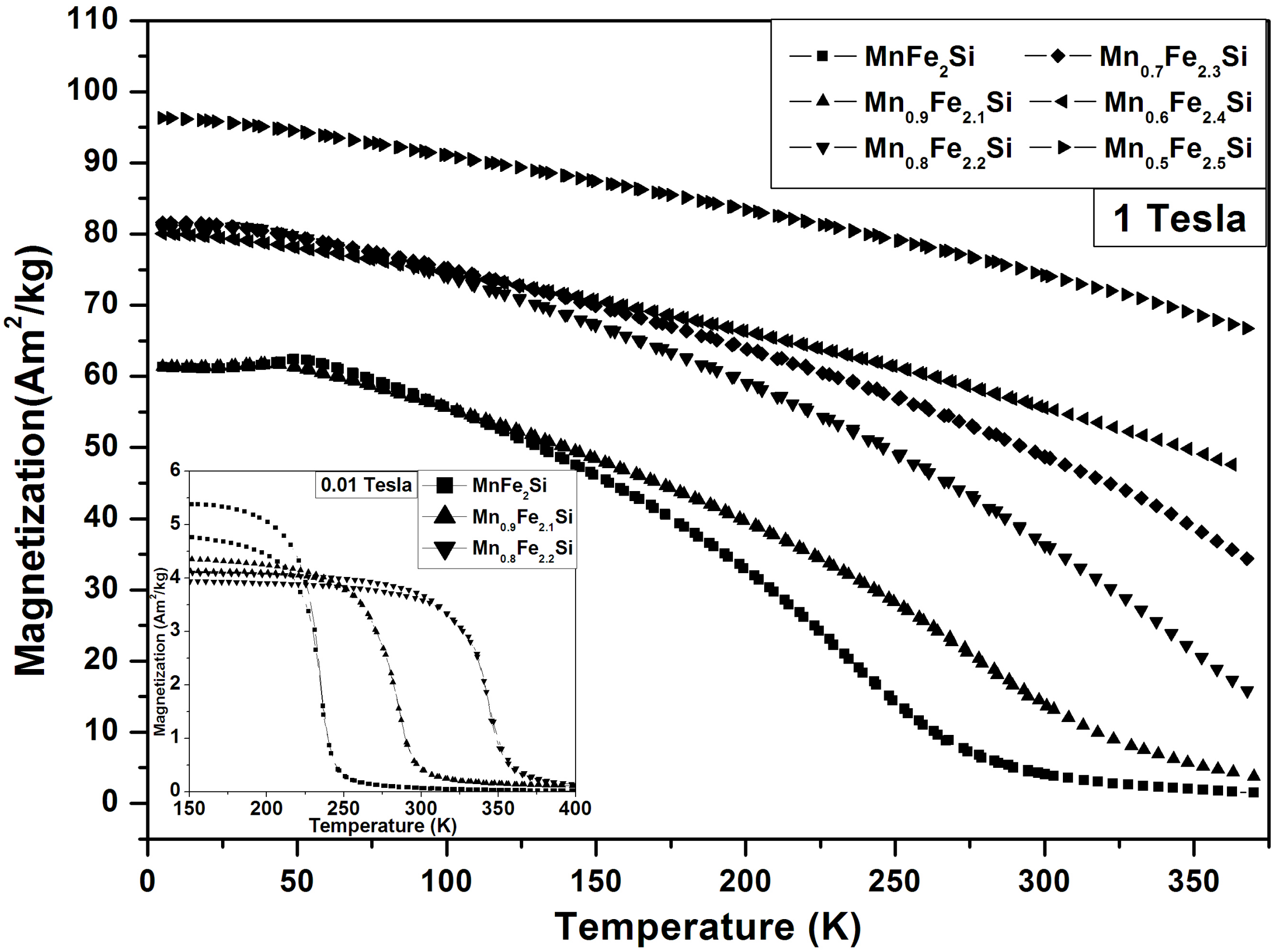

**Figure 5 - Magnetization versus temperature plot for $Mn_{3-x}Fe_xSi$, with x between 2 and 2.5, revealing the possible effects of the site preference of Mn and Fe atoms described in section 2.1 in magnetization. Insert: magnetization versus temperature measurements for low magnetic fields (0.01 T), revealing the change in $T_C$ with Fe content.**

## 5 – The $(Mn,Fe)_3(Si,P)$ magnetostructural phase diagram

By compiling all the data from our X-ray diffraction patterns, magnetization measurements and DSC measurements, together with data from [17], [19], [24], [25] and [26], we were thus able to construct a magnetostructural phase diagram of the $(Mn,Fe)_3(Si,P)$ system, Figure 6. Details regarding every sample produced for the current paper can be consulted in Table 1.

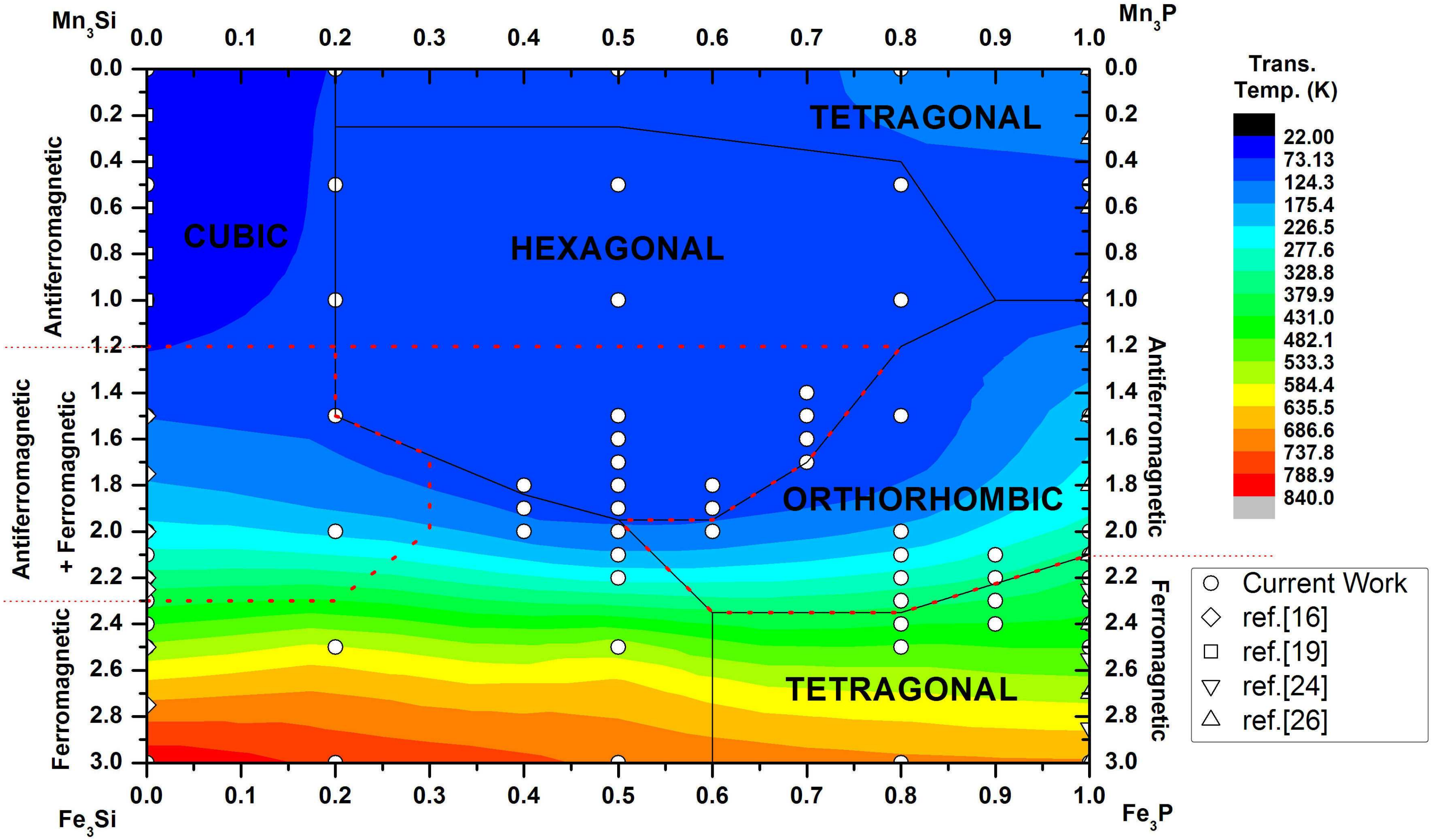

**Figure 6 - Magnetostructural phase diagram of the (Mn,Fe)$_3$(Si,P) system, revealing the compositional areas of all the different crystal structures described in the current paper and their magnetic behaviors at room temperature, for an annealing temperature of 950 ºC. The color code refers to transition temperatures ($T_N$ for the antiferromagnetic samples and $T_C$ for the ferromagnetic samples). In the canted ferromagnetic phase area (referred to in this diagram as the "antiferromagnetic + ferromagnetic" area), only the values of $T_C$ have been inserted in the diagram, seeing as $T_N$ here is relatively constant.**

## 5.1 - Overall diagram description

The $Mn_3Si$ compound orders in the cubic Fm3m structure and is an antiferromagnet below a $T_N$ of about 25 K, being paramagnetic above this temperature. This behavior is maintained for a P substitution up to 0.2, where we find the tetragonal $I\bar{4}$ structure. This structure is then observed until the full substitution of Si by P, in the $Mn_3P$ compound, as well as an increase in $T_N$ up to 115 K.

With increasing Fe substitution the cubic structure maintains its border along the $Mn_{3-x}Fe_xSi_{0.8}P_{0.2}$ line and displays a slow increase in $T_N$ until the Fe content of 1.2. The tetragonal phase, however, is no longer observed at an Fe substitution of only 0.5. Instead the hexagonal structure, consistent with the P6/mmm space group, is detected. This structure is only present in the P interval between 0.2 and 0.9, roughly, being bordered by the cubic phase of the Si rich side and by the tetragonal phase on the P rich side. It also exhibits antiferromagnetic behavior up to the Fe content of 1.2, much to the likeness of the cubic phase.

The $T_N$ of both the hexagonal and the tetragonal structures also demonstrates a slow increase with Fe content.

At an Fe content of 1, on the P rich side of the diagram, the orthorhombic structure is observed. This structure appears to be consistently antiferromagnetic with a relatively constant $T_N$ of 340 K, and is observed as far as the composition $Mn_{1.95}Fe_{1.05}Si_{0.5}P_{0.5}$.

At an Fe substitution of 1.2 both the cubic and hexagonal phases start to display ferromagnetic behavior. The $T_C$ of both these phases display an increase with Fe content, although the cubic phase does so at a much higher rate.

From the Fe content of 1.2 up to 2.3 the cubic phase, besides the transition between ferromagnetic and paramagnetic phases, also displays a transition between antiferromagentic and ferromagnetic phases, with a relatively constant $T_N$ of 50 K. This behavior is described as canted ferromagnetism.

At an Fe content of 1.9 the hexagonal phase is no longer observed, being substituted by the cubic phase whose range can now be observed from 0 to 0.5 P substitution. From the Fe content of 1.9 to 2.35 the cubic phase borders the orthorhombic phase.

As we further increase the Fe content $T_C$ of the cubic phase also continues to increase until it reaches its highest value of 840 K in the $Fe_3Si$ compound.

On the P rich side, the orthorhombic phase reaches its border at an Fe content of about 2.1, and the tetragonal $I\bar{4}$ structure is once again observed until the complete substitution of Mn by Fe. In this area the tetragonal phase now displays a transition between a ferromagnetic and a paramagnetic state, with an Fe dependent increase in $T_C$ until it reaches its maximum of 700 K in the $Fe_3P$ compound.

In the Fe rich region the cubic and tetragonal phases display a border along the $Mn_{3-x}Fe_xSi_{0.4}P_{0.6}$ with $2.35<x<3$.

All paramagnetic-ferromagnetic phase transitions observed in this system present the characteristics of second order phase transitions.

**Table 1 - Details regarding crystal structure, magnetic behavior and transition temperatures for every sample produced for the current paper**

| Sample Composition | Present Phases (Room Temp.) | Magnetic Transitions | Trans. Temp. |
|---|---|---|---|
| $Mn_3Si$ | Fm3m | Antiferro-Para | 27 K |
| $Mn_{2.5}Fe_{0.5}Si$ | Fm3m | Antiferro-Para | 24.5 K |
| $Mn_2FeSi$ | Fm3m | Antiferro-Para | 50 K |
| $Mn_{1.5}Fe_{1.5}Si$ | Fm3m | Antiferro-Ferro-Para | 62 \ 137 K |
| $MnFe_2Si$ | Fm3m | Antiferro-Ferro-Para | 50 \ 233 K |
| $Mn_{0.9}Fe_{2.1}Si$ | Fm3m | Antiferro-Ferro-Para | 40 \ 281 K |
| $Mn_{0.8}Fe_{2.2}Si$ | Fm3m | Antiferro-Ferro-Para | 30 \ 342 K |
| $Mn_{0.7}Fe_{2.3}Si$ | Fm3m | Antiferro-Ferro-Para | 12 \ 387 K |
| $Mn_{0.6}Fe_{2.4}Si$ | Fm3m | Ferro-Para | 397 K |
| $Mn_{0.5}Fe_{2.5}Si$ | Fm3m | Ferro-Para | 547 K |
| $Fe_3Si$ | Fm3m | Ferro-Para | 840 K |
| $Mn_3Si_{0.8}P_{0.2}$ | Fm3m+I$\bar{4}$ | Antiferro-Para | 79 K |
| $Mn_{2.5}Fe_{0.5}Si_{0.8}P_{0.2}$ | Fm3m+P6/mmm | Antiferro-Para | 82 K |
| $Mn_2FeSi_{0.8}P_{0.2}$ | Fm3m+P6/mmm | Antiferro-Para | 90 K |
| $Mn_{1.5}Fe_{1.5}Si_{0.8}P_{0.2}$ | Fm3m+P6/mmm | Antiferro-Ferro-Para | 58 \ 135 K |
| $MnFe_2Si_{0.8}P_{0.2}$ | Fm3m | Antiferro-Ferro-Para | 50 \ 173 K |
| $Mn_{0.5}Fe_{2.5}Si_{0.8}P_{0.2}$ | Fm3m | Ferro-Para | 657 K |
| $Fe_3Si_{0.8}P_{0.2}$ | Fm3m | Ferro-Para | 750 K |
| $Mn_{1.2}Fe_{1.8}Si_{0.6}P_{0.4}$ | P6/mmm+Fm3m | Ferro-Para | 102 K |
| $Mn_{1.1}Fe_{1.9}Si_{0.6}P_{0.4}$ | P6/mmm+Fm3m+Pmmm | Ferro-Para | 111 K |
| $MnFe_2Si_{0.6}P_{0.4}$ | Fm3m+P6/mmm+Pmmm | Ferro-Para | 111 K |
| $Mn_3Si_{0.5}P_{0.5}$ | I$\bar{4}$ | Antiferro-Para | 100 K |
| $Mn_{2.5}Fe_{0.5}Si_{0.5}P_{0.5}$ | P6/mmm | Antiferro-Para | 91 K |
| $Mn_2FeSi_{0.5}P_{0.5}$ | P6/mmm | Antiferro-Para | 107 K |
| $Mn_{1.5}Fe_{1.5}Si_{0.5}P_{0.5}$ | P6/mmm | Ferro-Para | 95 K |
| $Mn_{1.4}Fe_{1.6}Si_{0.5}P_{0.5}$ | P6/mmm | Ferro-Para | 85 K |
| $Mn_{1.3}Fe_{1.7}Si_{0.5}P_{0.5}$ | P6/mmm | Ferro-Para | 97 K |
| $Mn_{1.2}Fe_{1.8}Si_{0.5}P_{0.5}$ | P6/mmm+Pmmm | Ferro-Para | 108 K |
| $Mn_{1.1}Fe_{1.9}Si_{0.5}P_{0.5}$ | P6/mmm+Pmmm+Fm3m | Ferro-Para | 125 K |
| $MnFe_2Si_{0.5}P_{0.5}$ | Fm3m+P6/mmm+Pmmm | Ferro-Para | 127 K |
| $Mn_{0.9}Fe_{2.1}Si_{0.5}P_{0.5}$ | Fm3m+Pmmm+P6/mmm | Ferro-Para | 164 K |
| $Mn_{0.8}Fe_{2.2}Si_{0.5}P_{0.5}$ | Fm3m+Pmmm | Ferro-Para | 221 K |
| $Mn_{0.5}Fe_{2.5}Si_{0.5}P_{0.5}$ | Fm3m+I$\bar{4}$ | Ferro-Para | 667 K |
| $Fe_3Si_{0.5}P_{0.5}$ | Fm3m+I$\bar{4}$ | Ferro-Para | 697 K |
| $Mn_{1.2}Fe_{1.8}Si_{0.4}P_{0.6}$ | P6/mmm+Pmmm | Ferro-Para | 103 K |
| $Mn_{1.1}Fe_{1.9}Si_{0.4}P_{0.6}$ | P6/mmm+Pmmm | Ferro-Para | 127 K |
| $MnFe_2Si_{0.4}P_{0.6}$ | Pmmm+Fm3m | Antiferro-Para | 84 K |
| $Mn_{1.6}Fe_{1.4}Si_{0.3}P_{0.7}$ | P6/mmm+Pmmm | Ferro-Para | 71 K |
| $Mn_{1.5}Fe_{1.5}Si_{0.3}P_{0.7}$ | P6/mmm+Pmmm | Ferro-Para | 75 K |
| $Mn_{1.4}Fe_{1.7}Si_{0.3}P_{0.7}$ | Pmmm+P6/mmm | Ferro-Para | 84 K |
| $Mn_{1.3}Fe_{1.7}Si_{0.3}P_{0.7}$ | Pmmm+P6/mmm | Ferro-Para | 88 K |
| $Mn_3Si_{0.2}P_{0.8}$ | $I\bar{4}$ | Antiferro-Para | 133 K |
| $Mn_{2.5}Fe_{0.5}Si_{0.2}P_{0.8}$ | P6/mmm+$I\bar{4}$ | Antiferro-Para | 117 K |
| $Mn_2FeSi_{0.2}P_{0.8}$ | P6/mmm | Antiferro-Para | 133 K |
| $Mn_{1.5}Fe_{1.5}Si_{0.2}P_{0.8}$ | Pmmm | Antiferro-Para | 78 K |
| $MnFe_2Si_{0.2}P_{0.8}$ | Pmmm+$I\bar{4}$ | Antiferro-Para | 154 K |
| $Mn_{0.9}Fe_{2.1}Si_{0.2}P_{0.8}$ | Pmmm+$I\bar{4}$ | Antiferro-Para | 233 K |
| $Mn_{0.8}Fe_{2.2}Si_{0.2}P_{0.8}$ | Pmmm+$I\bar{4}$ | Antiferro-Para | 223 K |
| $Mn_{0.7}Fe_{2.3}Si_{0.2}P_{0.8}$ | Pmmm+$I\bar{4}$ | Antiferro-Para | 213 K |
| $Mn_{0.6}Fe_{2.4}Si_{0.2}P_{0.8}$ | $I\bar{4}$ +Pmmm | Ferro-Para | - |
| $Mn_{0.5}Fe_{2.5}Si_{0.2}P_{0.8}$ | $I\bar{4}$ +Pmmm | Ferro-Para | 523 K |
| $Fe_3Si_{0.2}P_{0.8}$ | $I\bar{4}$ | Ferro-Para | 695 K |
| $Mn_{0.9}Fe_{2.1}Si_{0.1}P_{0.9}$ | Pmmm+$I\bar{4}$ | Antiferro-Para | 227 K |
| $Mn_{0.8}Fe_{2.2}Si_{0.1}P_{0.9}$ | Pmmm+$I\bar{4}$ | Antiferro-Para | 219 K |
| $Mn_{0.7}Fe_{2.3}Si_{0.1}P_{0.9}$ | $I\bar{4}$ +Pmmm | Ferro-Para | - |
| $Mn_{0.6}Fe_{2.4}Si_{0.1}P_{0.9}$ | $I\bar{4}$ +Pmmm | Ferro-Para | - |
| $Mn_3P$ | $I\bar{4}$ | Antiferro-Para | 150 K |
| $Mn_{2.5}Fe_{0.5}P$ | $I\bar{4}$ | Antiferro-Para | 117 K |
| $Mn_2FeP$ | $I\bar{4}$ +Pmmm | Antiferro-Para | 89 K |
| $Mn_{1.5}Fe_{1.5}P$ | Pmmm | Antiferro-Para | 233 K |
| $MnFe_2P$ | Pmmm+$I\bar{4}$ | Antiferro-Para | 240 K |
| $Mn_{0.9}Fe_{2.1}P$ | Pmmm+$I\bar{4}$ | Ferro-Para | 400 K |
| $Mn_{0.8}Fe_{2.2}P$ | Pmmm+$I\bar{4}$ | Ferro-Para | 400 K |
| $Mn_{0.7}Fe_{2.3}P$ | $I\bar{4}$ +Pmmm | Ferro-Para | 413 K |
| $Mn_{0.6}Fe_{2.4}P$ | $I\bar{4}$ +Pmmm | Ferro-Para | 420 K |
| $Mn_{0.5}Fe_{2.5}P$ | $I\bar{4}$ +Pmmm | Ferro-Para | 473 K |
| $Fe_3P$ | $I\bar{4}$ | Ferro-Para | 700 K |

## 6 - Conclusion

Based on the collected data and previous literature, the magnetostructural quaternary phase diagram of the $(Mn,Fe)_3(Si,P)$ system was successfully constructed, offering a rare overall view of its unique magnetic and structural properties and greatly expanding our knowledge of the already widely studied $(Mn,Fe)_3Si$ and $(Mn,Fe)_3P$ systems. Still, this diagram appears to be very particular and dependent to our sample production procedure, as other authors have found certain phases we have not [17] and we have also observed a dependence of phase percentages near border areas with the annealing temperature.

The novel hexagonal phase, observed for the first time in this system, having both paramagnetic and ferromagnetic behaviors, may offer the possibility for further developments on those areas where both the $(Mn,Fe)_3Si$ and $(Mn,Fe)_3P$ systems have been studied in the past, such as spin-wave excitation by means of neutron inelastic scattering [8], in its paramagnetic phase, spintronics research [9], in its ferromagnetic phase or on metal–metalloid compound research, an area of interest in nuclear-reactor materials science [21].
Even though we have found this system inappropriate, by itself, for magnetocaloric applications, this research and the understanding it offers may still open the possibility for further magnetocaloric studies and developments. The novel hexagonal phase still represents a new and unexplored set of compounds that may still be optimized, either by the addition of a fifth element or the insertion of interstitial atoms into this structure. Also, based on the knowledge of structural phase borders, these same techniques, and fine-tuning of sample production procedures, may still be employed in an attempt to trigger temperature and\or magnetic field dependent structural transitions with great magnetocaloric potential.

## 7 - Funding and Acknowledgment

The authors wish to acknowledge the funding from BASF Future Business and FOM (Stichting voor Fundamenteel Onderzoek der Materie), under the Industrial Partnership Programme IPP I18 of the 'Stichting voor Fundamenteel Onderzoek der Materie (FOM)' which is financially supported by the 'Nederlandse Organisatie voor Wetenschappelijk Onderzoek (NWO)', to the accomplishment of the research presented in this paper.

## 8 - References